\documentclass[12pt]{article}
\usepackage[cp866]{inputenc}
\usepackage{hyperref,amsfonts,amssymb,theorem}

\def\bq{ \begin{equation} }
\def\eq{ \end{equation} }
\def\ben{ \begin{eqnarray} }
\def\en{ \end{eqnarray} }

\def\R{\mathbb{R}}

\newtheorem{thr}{Theorem}[section]
\theorembodyfont{\sl}
\newtheorem{prop}{Proposition}[section]
\theorembodyfont{\rm}
\newtheorem{ex}{Example}[section]
\newtheorem{rem}{Remark}[section]

\begin{document}

\title{\bf Factorization of the current algebra and integrable top-like systems.}

\author{I.Z. Golubchik\\
\\Bashkirian State Pedagogical University,\\
October Revolution str. 3 a, Ufa 450000, Russia\\ \\
V.V. Sokolov\\ \\Landau Institute for Theoretical Physics,
\\ Kosygina str. 2, Moscow, 119334,
Russia\\ }

\date{}
\maketitle

\begin{abstract} A hierarchy of integrable hamiltonian nonlinear ODEs is associated with
any decomposition of the Lie algebra of Laurent series with coefficients being elements of
a semi-simple Lie algebra into a sum of the subalgebra consisting of the Taylor series and
some complementary subalgebra. In the case of the Lie algebra $so(3)$ our scheme covers
all classical integrable cases in the Kirchhoff problem of the motion of a
rigid body in an ideal fluid.  Moreover, the construction allows us to generate integrable
deformations for known integrable models.

\end{abstract}


\hskip3cm

\section{Introduction}
\setcounter{equation}{0}

In the paper we consider integrable nonlinear ODEs with Lax representations defined
by a vector space decomposition of the algebra ${\cal G_{\lambda}}$ of Laurent series
with respect to the parameter
$\lambda$ with coefficients being elements of a semi-simple Lie algebra ${\cal G}$ into a
direct sum
\begin{equation}\label{decomp}
{\cal G}_{\lambda}={\cal G}_{+}\oplus {\cal G}_{-},
\end{equation}
where ${\cal G}_{-}$ is the subalgebra of Taylor series and ${\cal
G}_{+}$ is some complementary subalgebra. Following the
terminology by I. Cherednik \cite{chered}, we shall call such complementary subalgebras
{\it factorizing} ones. For the standard situation,  ${\cal G}_{+}$ coincides
with the set ${\cal G}_{+}^{st}$ of all polynomials in $\lambda^{-1}$
with zero free term.

In the case when the factorizing subalgebra is isotropic with respect to the
invariant nondegenerate form
\begin{equation} \label{forma}
<X(\lambda),\, Y(\lambda)> = {\rm res} \Big(X(\lambda),\,Y(\lambda)\Big), \qquad
X(\lambda),Y(\lambda) \in {\cal G_{\lambda}},
\end{equation}
where $(\cdot,\cdot)$ is the Killing form on
${\cal G}$, there exist  deep interconnections between factorizing subalgebras and solutions
of the classical Yang-Baxter equation \cite{beldrin}. Non-isotropic subalgebras probably have not been
seriously investigated.

Equations of the Landau-Lifshitz type form the most known class of integrable PDEs related
to the factorizing subalgebras \cite{golsok1}. The Lax representation for these equations has the form
\begin{equation}\label{laxpde}
L_{t}=A_{x}+[A,\, L],
\end{equation}
where $L, A$ are elements of a factorizing algebra ${\cal G}_{+}$. A relationship between
factorizing subalgebras and systems of the chiral model type has been established in
\cite{golsok2}.

In the case if a solution of the equation of the Landau-Lifshitz type does not depend on $x$,
relation (\ref{laxpde}) is reduced to
\begin{equation}\label{laxode}
L_{t}= [A,\, L].
\end{equation}
While  for equation (\ref{laxpde}) it is naturally to suppose that both
$L$ and $A$ belong to the same Lie algebra, equation
(\ref{laxode}) is self-consistent also in the case when $A$ belongs to a Lie
algebra whereas $L$ lies in some module over this algebra. One of our main
observations is that besides equations
(\ref{laxode}), where  $L, A \in {\cal G}_{+}$, there is a second, maybe more interesting,
class of integrable nonlinear ODEs, for which
$A \in {\cal G}_{+}$, $L \in {\cal G}_{+}^{\perp}.$
Here $\perp$ stands for the orthogonal complement with respect to form
(\ref{forma}). These two classes coincide for isotropic factorizing
subalgebras. It is easy to see that ${\cal G}_{+}^{\perp}$ is a module
over ${\cal G}_{+}.$  It is especially important for us that equations on ${\cal
G}_{+}^{\perp}$ possesses a natural hamiltonian structure defined by a
linear Poisson bracket. In general, equations of the second class do not admit any two-dimensional
generalization (\ref{laxpde}).

Apparently, the Lax equations with $L \in {\cal G}_{+}^{\perp}$ slip expert's minds.
It is shown in Section 4 that precisely such Lax representations for ${\cal G}=so(3)$ correspond
to classical integrable homogeneous quadratic Hamiltonians in rigid body
dynamics (namely, to spinning tops by Euler, Kirchhoff,  Clebsch and
Steklov-Lyapunov). All factorizing subalgebras for ${\cal G}=so(3)$ have
been classified in the paper \cite{sokdan}. It turns out that the classical tops
are in one-to-one correspondence with the factorizing subalgebras.
In addition, various deformations of known integrable models found recently in
\cite{soktmf,sok1,sokts1} turn out to be generated by non-isotropic factorizing subalgebras
(see Section 5).

In Sections 2,3 a general algebraic theory for integrable models related to
factorizing subalgebras is developed. In particular, a description of
commuting flows and linear Poisson brackets on finite-dimensional orbits is
given. It is interesting to note that if the factorizing subalgebra
satisfies the following condition
\begin{equation}\label{genhom1}
\lambda^{-m} {\cal G}_{+}\subset {\cal G}_{+}
\end{equation}
for some $m\ge 1$, then the equations of motion are hamiltonian with respect to
one more linear Poisson bracket. This fact explains the existence of linear
transformations found by A. Bobenko \cite{bobenko}, which
interconnect known integrable cases on $e(3)$ and $so(4).$ The compatibility of
two linear Poisson brackets has been taking as a basis for the approach
developed in \cite{borbol} (see also \cite{golsok2}).

If condition (\ref{genhom1}) is fulfilled with $m=1$,
the factorizing subalgebra ${\cal G}_{+}$ is said to be  {\it homogeneous}.
In Section 6 the homogeneous subalgebras are considered. In particular, we
show that the factorizing subalgebra associated with any finite group of reduction
(see \cite{mikhailov1} ) is equivalent to a homogeneous one.

{\bf Acknowledgements.} The authors are grateful to V.G.~Drinfeld,
A.V.~Mikhailov, and M.A.~Semenov-Tian-Shansky for useful discussions.
The research was partially supported by RFBR grants 02-01-00431 and
NSh 1716.2003.1.

\section{General scheme.}
Let ${\cal A}$ be an associative algebra equipped with nondegenerate
semi-Frobenius bilinear form
\linebreak $<\cdot,\cdot>$. The latter means that
$$
<a,b>=<b,a>, \qquad <a, b\star c>=<a\star b,c>.
$$
By ${\cal A}^{(-)}$ denote the adjoint Lie algebra with the bracket
$[a,b]=a\star b-b \star a.$ It is easy to see that $<\cdot,\cdot>$ induces
the following invariant form on ${\cal A}^{(-)}$. In other words,
\begin{equation}\label{inv}
<[a,b], c>=-<b, [a,c]>.
\end{equation}

Let ${\cal U}$ be a Lie subalgebra in ${\cal A}^{(-)}$ such that the restriction
$<\cdot,\cdot>_{{\cal U}}$ of the form $<\cdot,\cdot>$ to ${\cal U}$ is nondegenerate and
\begin{equation}\label{ort}
{\cal A}^{(-)}={\cal U}\oplus {\cal U}^{\perp}.
\end{equation}
Suppose that
\begin{equation}\label{dirsum}
{\cal U}={\cal U}_{+}\oplus {\cal U}_{-},
\end{equation}
where ${\cal U}_{-}$ and ${\cal U}_{+}$ are Lie subalgebras. By
$\pi_{-}$ and $\pi_{+}$ denote the projectors onto ${\cal U}_{-}$ and ${\cal U}_{+}$,
respectively.

Let ${I}_{-}$ and ${I}_{+}$ be ideals of finite codimension in ${\cal U}_{-}$
and ${\cal U}_{+}$. Consider a vector subspace in ${\cal U}$ of the form
\begin{equation}\label{genort}
{\cal O}=I_{-}^{\perp}\cap I_{+}^{\perp},
\end{equation}
where ${\perp}$ stands for the orthogonal complement in ${\cal U}$ with respect to the form
$<\cdot,\cdot>_{{\cal U}}$. Since
$$
I_{-}^{\perp}\cap I_{+}^{\perp}=(I_{-} + I_{+})^{\perp},
$$
the vector space ${\cal O}$ is finite-dimensional.

\begin{ex}\label{rrr} Let ${\cal A}$ be arbitrary associative algebra,
${\cal U}$ be the Lie algebra of Laurent series with respect to the parameter $\lambda$
with coefficients from ${\cal A}$, and $r$ be a fixed constant element of ${\cal A}$. It is easy to see that
\begin{equation}\label{comp}
{\cal U}_{+}=\{ \sum^{p}_{i=1} a_{i} \lambda^{-i} (1+\lambda r)
\quad \vert \quad a_{i}\in {\cal A}, \quad p\in {\mathbb N} \}
\end{equation}
is a factorizing Lie subalgebra in ${\cal U}$, which is complementary to the subalgebra
${\cal U}_{-}$ of all Taylor series. One can verify that
\begin{equation}\label{ortcomp}
{\cal U}_{+}^{\perp}=\{ \sum^{q}_{i=1} (1+\lambda r)^{-1} a_{i}
\lambda^{-i} \quad \vert \quad a_{i}\in {\cal A}, \quad q\in
{\mathbb N} \}.
\end{equation}
Define the orbit ${\cal O}_n$ by the following pair of ideals
$$
I_{+}={\cal U}_{+}, \qquad I_{-}=\lambda^n {\cal U}_{-}.
$$
It follows from (\ref{genort}) that  $$ {\cal O}_{n}=\lambda^{-n}{\cal U}_{-}\cap {\cal
U}_{+}^{\perp}. $$ Thus we have $$ {\cal O}_{n}=\{
\sum^{n}_{i=1} (1+\lambda r)^{-1} a_{i} \lambda^{-i} \quad \vert
\quad a_{i}\in {\cal A} \}. $$
\end{ex}

Let  $f(L,\lambda_1,\cdots \lambda_k)$ be a polynomial with constant
coefficients such that $L\in {\cal O}$ and $\lambda_i$ are arbitrary elements from the center of  ${\cal A}$.
By $\bar f$ denote the projection of $f$ onto ${\cal U}$ corresponding
to the decomposition (\ref{ort}). Expressions like $\bar f$ we shall call {\it spectral invariants}
of $L$. Since $[f,\,L]=0,$ we have also  $[\bar f,\,L]=0.$
If $\bar g$ is one more spectral invariant, then
$[\bar f,\,L]=0$ implies $[\bar f,\,g]=0.$ Now it follows from
(\ref{ort}) that  $[\bar f,\,\bar g]=0.$

\begin{thr} The set ${\cal O}$ is a finite-dimensional orbit with respect to
the flow defined by the Lax equation
\begin{equation}\label{genlaxf}
L_{t}=[\pi_{+}(\bar f),\, L],
\end{equation}
where $\bar f$ is arbitrary spectral invariant of $L$.
\end{thr}

\par\noindent\textbf{Proof.} In order to prove that ${\cal O}$ is an orbit,
we have to verify that  $L \in {\cal O}$ implies
$[\pi_{+}(\bar f ),\, L] \in {\cal O}$. By definition,
$\pi_{+}(\bar f)\in {\cal U}_{+}$. It follows from the invariance of
$<\cdot,\cdot>$ that $I_{+}^{\perp}$ is a module over ${\cal U}_{+}$.
Therefore $[\pi_{+}(\bar f),\, L] \in
I_{+}^{\perp}$. From $[\bar f,\, L]=0$, we get $[\pi_{+}(\bar
f),\, L]=-[\pi_{-}(\bar f),\, L]$. As $I_{-}^{\perp}$ is a module over ${\cal U}_{-}$,
we have $[\pi_{+}(\bar f),\, L]\in
I_{-}^{\perp}$. This implies that $[\pi_{+}(\bar f),\, L]\in {\cal O}$.

As we will see below, the Lax equation possesses a rich store of first
integrals and commuting flows.

\begin{prop} Expression $<\bar g_1,\bar g_2>$ is a first integral of equation (\ref{genlaxf}) for any two
spectral invariants $h=\bar g_1$ and $\bar g_2.$
\end{prop}

\par\noindent\textbf{Proof.} Since $g_i$ is a polynomial in $L$, we have
$(g_i)_t=[\pi_{+}(\bar f),\, g_i]$. It follows from the decomposition (\ref{ort}) that
\begin{equation}\label{gplus}
(\bar g_i)_t=[\pi_{+}(\bar f),\, \bar g_i].
\end{equation}
The invariance  (\ref{inv}) of the form $<\cdot,\cdot>$ guarantees that the
expression
$$h_t=<\bar g_1, [\pi_{+}(\bar f),\, \bar g_2]>+<\bar g_2,
[\pi_{+}(\bar f),\, \bar g_1]>.
$$
is equal to zero.

\begin{thr} For any spectral invariants  $\bar g_1$ and $\bar g_2,$
the flows
\begin{equation}\label{twolax1}
L_{t}=[\pi_{+}(\bar  g_1),\, L]
\end{equation}
and
\begin{equation}\label{twolax2}
 L_{\tau}=[\pi_{+}(\bar g_2),\, L]
\end{equation}
on the orbit ${\cal O}$ commute each other. \end{thr}

\par\noindent\textbf{Proof.} Using (\ref{gplus}),
(\ref{twolax1}), (\ref{twolax2}), we get
$$
L_{t \tau}-L_{\tau t}=\Big[\pi_{+} [\pi_{+}(\bar g_2),\, \bar g_1]- \pi_{+}
[\pi_{+}(\bar g_1),\, \bar g_2]-[\pi_{+}(\bar g_2), \pi_{+}(\bar g_1)], \,
L\Big].
$$
Since both summands in decomposition (\ref{dirsum}) are supposed to be
subalgebras, the latter equality can be written as follows
\begin{equation}\label{tw1}
L_{t \tau}-L_{\tau t}=\Big[\pi_{+} [\bar g_2,\, \bar g_1], \, L\Big].
\end{equation}
The right hand side of this relation vanishes because, as it was mentioned
above, $[\bar g_2,\, \bar g_1]=0.$

It turns out that under the following natural (cf. (\ref{genort})) additional condition
$$
{\cal O}^{\perp}=I_{-}+I_{+},
$$
equation (\ref{genlaxf}) is a hamiltonian one. Define ${\cal H}$ by the
formula
$$
{\cal H}={\cal U}_{-}/I_{-}\oplus \bar{\cal U}_{+}/I_{+},
$$
where the algebra $\bar{\cal U}_{+}$ is obtained from ${\cal U}_{+}$ by the replacement
of the bracket
$[a,b]$ by $[b,a]$. Stress that  the symbol $\oplus$ in this formula stands
for the direct sum of Lie algebras in
contrast with (\ref{dirsum}), where it means the vector space direct sum.

Define a nondegenerate bilinear form, pairing the algebra ${\cal
H}$ and the orbit ${\cal O}$ as follows
\begin{equation}\label{form3}
\Big((a_{-}+I_{-},a_{+}+I_{+}),L\Big)=<a_{+}+a_{-}, L>.
\end{equation}
Since $<I_{-}\oplus I_{+}, {\cal O}>=0,$ the form (\ref{form3}) is well defined.
The Kirillov bracket on ${\cal O}$ is given by
\begin{equation}\label{Kirilgen}
\{\varphi,\, \psi  \}(L)=([\mbox{\rm grad}_L \varphi, \mbox{\rm
grad}_L \psi ],\, L), \qquad L\in {\cal O}, \quad \mbox{\rm
grad}_L \varphi, \mbox{\rm grad}_L \psi \in {\cal H}.
\end{equation}
Suppose that the bases $f^i$ and $e_i$ in ${\cal O}$ and ${\cal H}$ are adjoint with respect to
the form (\ref{form3}), $$ [e_i,\,
e_j]=c_{ij}^k e_k$$ and $L=q_i f^{i}$. Then, because of the formula $\mbox{\rm
grad}\, q_i=e_i,$ the Poisson brackets between the coordinate functions are given
by
\begin{equation} \label{puasson}
\{q_i,\, q_j\}=c_{ij}^k q_k.
\end{equation}

\begin{thr} (cf. \cite{reysem})
\begin{itemize}
\item i) \,  The functionals of the form
\begin{equation} \label{func}
H_g=<g,\, 1>,
\end{equation}
where $g(L,\lambda_1,\cdots \lambda_k)$ is a polynomial in all arguments with constant
coefficients and $L\in {\cal O}$, are in involution with respect to the bracket (\ref{puasson});
\item ii) \, The Lax equation
\begin{equation} \label{eeq}
L_{t}=[\pi_{+}(\bar f),\, L], \
\end{equation}
where
\begin{equation} \label{gf}
f = -\frac{d g}{d L} \Big(L,\lambda_1,\cdots \lambda_k)\Big),
\end{equation}
on the orbit ${\cal O}$ is hamiltonian  with respect to the bracket
(\ref{puasson}) and the Hamiltonian function $ H_g$.
\end{itemize}
\end{thr}

\par\noindent\textbf{Proof.} Let $H_{g_{1}}$ and
$H_{g_{2}}$ be functionals of the form (\ref{func}). Then
$$
\mbox{\rm grad} H_{g_{i}}= - \Big(\pi_{-}(\bar f_i)+I, \,
\pi_{+}(\bar f_i)+J\Big),
$$
where $f_i$ and $g_i$ are related by the formula (\ref{gf}).
According to (\ref{Kirilgen}), we have $$
\{H_{g_{1}},H_{g_{2}}\}=<-[\pi_{+}(\bar f_1),\pi_{+}(\bar f_2)]+
[\pi_{-}(\bar f_1),\pi_{-}(\bar f_2)],\, L>= $$ $$
\frac{1}{2}<[(-\pi_{+}(\bar f_1)+\pi_{-}(\bar f_1)),\, \bar f_2]-
[(-\pi_{+}(\bar f_2)+\pi_{-}(\bar f_2)),\, \bar f_1],\, L
>.
$$
It follows from the invariance of the form $<\cdot,\cdot>$ and from
the relations $[\bar f_1,\, L]=[\bar f_2,\,L]=0,$ that the latter expression is equal to zero.
The item 1 is proved. The hamiltonian system with the Hamiltonian $H_g$
has the form $$ \frac{d \Psi(L)}{dt}=\{\Psi,\,
H_g\}(L)=(\mbox{\rm grad}_L(\Psi), \,L_t). $$ From the other hand,
$\mbox{\rm grad}_L(\Psi)=(\pi_{-}(a)+I_{-},\pi_{+}(a)+I_{+})$ for some
$a\in{\cal U}$. Therefore, $$ (\mbox{\rm grad}_L(\Psi),
[\pi_+(\bar
f),\,L])=\frac{1}{2}\Big((\pi_{-}(a)+I_{-},\pi_{+}(a)+I_{+}),
[\pi_{+}(\bar f)-\pi_{-}(\bar f), \, L]\Big)= $$ $$
\frac{1}{2}<[\pi_{-}(a)+\pi_{+}(a),\pi_{+}(\bar f)-\pi_{-}(\bar
f)]+[\pi_{-}(a)-\pi_{+}(a),\bar f ],\,L>= $$ $$
<[\pi_{+}(a),\pi_{+}(\bar f)]-[\pi_{-}(a),\pi_{-}(\bar f)],\,L >=
$$ $$ \Big([\mbox{\rm grad}_L(\Psi),\,\mbox{\rm grad}_L(H_g)], \,L \Big)=\{\Psi,\,H_g
\}(L). $$
In the next to last equality we used the formula
$$\mbox{\rm grad} H_g=-(\pi_{-}(\bar f)+I_{-},\pi_{+}(\bar f)+I_{+}),$$
where $g$ and $f$ are related by (\ref{gf}). The nondegeneracy of the form $(\cdot,\cdot)$ implies that
the equation  $L_t=[\pi_+(\bar f),\,L]$ is hamiltonian with respect to the
bracket (\ref{Kirilgen}). This completes the proof of Theorem.

Suppose the ideals $I_{+}$ and $I_{-}$ satisfy the following additional
conditions:
\begin{equation} \label{genhom}
\mu \,I_{+}\subset {\cal G}_{+}, \qquad \mu \,
I_{-}\subset {\cal G}_{-},\qquad \mu^{-1} {\cal O }\subset {\cal U}
\end{equation}
for some invertible element $\mu$ from the center of the associative algebra ${\cal A}$.
Then:

\begin{rem} The orbit ${\cal O}_{\mu}$ constructed by means of the ideals
$\mu \,I_{+}$ and $\mu \, I_{-}$,
is related to ${\cal O}$ by the formula $$ {\cal
O}_{\mu}=\mu^{-1}\,{\cal O}.
$$
\end{rem}

\begin{rem}
The equation $ \tilde L_{t}=[\pi_{+}(\bar F(\tilde
L,,\lambda_1,\cdots \lambda_k)),\, \tilde L]$ on the orbit ${\cal
O}_{\mu}$ coincides with equation (\ref{eeq}) on the orbit ${\cal
O}$, where $\tilde L=\mu^{-1} L$, $f(L,\lambda_1,\cdots
\lambda_k)=F(\tilde L,,\lambda_1,\cdots \lambda_k)$.
\end{rem}

\begin{rem}\label{rem3} Under condition (\ref{genhom}), Lax equation
(\ref{eeq}) has two hamiltonian structures described in Theorem
3. The first structure corresponds to the pair of ideals $I_{+},\,I_{-}$
whereas the second is related to the pair $\mu \,I_{+},\mu \,I_{-}$.
The question whether these Poisson brackets
are compatible remains still open.
\end{rem}

\section{The case of the current algebra.}
\setcounter{equation}{0}

In this paper we consider the current algebra over a semi-simple Lie algebra ${\cal G}$ as
a basic example of the Lie algebra ${\cal U}.$ We assume that ${\cal G}$ is embedded into the matrix
algebra $Mat_{k\times k}$ by the adjoint representation.
In this case the associative algebra ${\cal A}$ is the set of all Laurent series with respect to
the parameter $\lambda$ with coefficients from $Mat_{k\times k}$. The bilinear form on ${\cal A}$
is defined by the formula
\begin{equation}\label{formass}
<X(\lambda),\, Y(\lambda)>=\hbox{res}\Big( \hbox{trace} (X
Y)\Big).
\end{equation}
We take
\begin{equation}
{\cal G_{\lambda}}=\{ \sum^{\infty}_{i=-n} g_{i} \lambda^{i}\quad
\vert \quad g_{i}\in {\cal G}, \quad n\in {\mathbb Z}  \}
\end{equation}
and
\begin{equation}\label{taylor}
{\cal G}_{-}=\{ \sum^{\infty}_{i=0} g_{i} \lambda^{i}\quad \vert
\quad g_{i}\in {\cal G} \}
\end{equation}
for ${\cal U}$ and ${\cal U}_{-}$.
Suppose that as a vector space, ${\cal
G}_{\lambda}$ is a direct sum of ${\cal G}_{-}$ and some complementary factorizing
subalgebra ${\cal G}_{+}$.

The standard factorizing subalgebra is given by
\begin{equation}\label{polyn}
{\cal G}_{+}^{st}=\{ \sum^{n}_{i=1} g_{i} \lambda^{-i}\quad \vert
\quad g_{i}\in {\cal G}, \quad n \in {\mathbb N} \}.
\end{equation}
It is easy to see that both ${\cal G}_{-}$ and ${\cal G}_{+}^{st}$ are
isotropic with respect to the form
\begin{equation}\label{formad}
<X,\, Y>=\hbox{res}\Big( \hbox{trace} (ad_{X}\cdot
ad_{Y})\Big), \qquad X,Y\in {\cal G}_{\lambda}.
\end{equation}

Define the orbit ${\cal O}_n$ by the pair of ideals
\begin{equation}\label{ideal}
I_{+}={\cal G}_{+}, \qquad I_{-}=\lambda^n {\cal G}_{-}.
\end{equation}
It follows from (\ref{genort}) that
\begin{equation}\label{orbnst}
{\cal O}_{n}=\lambda^{-n}{\cal G}_{-}\cap {\cal G}_{+}^{\perp}.
\end{equation}
For the standard factorizing subalgebra   ${\cal
G}_{+}^{st}$ we have $$ {\cal O}_{n}=\{ \sum^{n}_{i=1} g_{i}
\lambda^{-i}\}. $$

The spectral invariants are supposed to be polynomials of the form
$f(L,\lambda_1,\lambda_{2})$, where $\lambda_1=\lambda,\quad
\lambda_2=\lambda^{-1}.$

The corresponding linear Poisson bracket (\ref{puasson}) is related to the
finite-dimensional Lie algebra
\begin{equation}\label{algebra1}
{\cal H}={\cal G}_{-}/\lambda^n {\cal G}_{-}.
\end{equation}
Since ${\cal G}_{-}$ coincides with the algebra of Taylor series,
the algebra ${\cal H}$ is isomorphic to the polynomial algebra
${\cal G}[\varepsilon]$, where $\varepsilon^n=0$. In particular, if
${\cal G}=so(3)$ and $n=2$, then ${\cal H}$ is isomorphic to the Lie algebra
$e(3)$ of motions of Euclidean space $\R^3.$

Remark \ref{rem3} states that if the subalgebra
${\cal G}_{+}$ satisfies condition (\ref{genhom1}) for some $0\le m\le n,$ then
the corresponding linear Poisson bracket
$\{\cdot,\cdot\}_m$ can be constructed such that the spectral invariants of the operator $L$
are in involution with respect to this bracket. The Poisson bracket $\{\cdot,\cdot\}_m$ corresponds to
the finite-dimensional Lie algebra
\begin{equation}\label{algebra2}
 {\cal H}_m={\cal G}_{-}/I\oplus \bar{\cal G}_{+}/J,
\end{equation}
where $I=\lambda^{n-m} {\cal G}_{-}, \quad J=\lambda^{-m} \bar {\cal
G}_{+}$. The algebra  $\bar{\cal G}_{+}$ is obtained from ${\cal G}_{+}$ by the replacement of $[a,b]$ with
$[b,a].$ The symbol $\oplus$ in (\ref{algebra2}) means the direct sum of Lie algebras.
If ${\cal G}=so(3),$ $n=2$, and $m=1$, then in generic case ${\cal H}_1$ is isomorphic to the
Lie algebra $so(4)$.

When $m=0,$ condition (\ref{genhom1}) is always fulfilled
and we find ourselves in the situation described by (\ref{algebra1}).
A subalgebra is called {\it homogeneous} if condition (\ref{genhom1}) with $m=1$ holds.
Section 6 is devoted to homogeneous subalgebras.

\section{The current algebra over $so(3)$ and classical spinning tops.}
\setcounter{equation}{0}

In this section we consider the case ${\cal G_{\lambda}}=so(3).$ Let us fix
the basis
$$
{\bf e_{1}}=\pmatrix{0&0&1 \cr 0&0&0 \cr -1&0&0\cr},
\quad {\bf e_{2}}=\pmatrix{0&0&0 \cr 0&0&1 \cr
0&-1&0\cr}, \quad {\bf e_{3}}=\pmatrix{0&1&0 \cr -1&0&0 \cr 0&0&0\cr}
$$
in the Lie algebra $so(3)$. It is easy to check that
$$
[{\bf e_{2}},\,{\bf e_{1}}]={\bf e_{3}},
\qquad [{\bf e_{3}},\,{\bf e_{2}}]={\bf e_{1}},
\qquad [{\bf e_{1}},\,{\bf e_{3}}]={\bf e_{2}}.
$$
Define the invariant bilinear form on $so(3)$ as
$$
(x_{1}{\bf e_{1}}+x_{2}{\bf e_{2}}+x_{3}{\bf e_{3}},
\quad y_{1}{\bf e_{1}}+y_{2}{\bf e_{2}}+y_{3}{\bf e_{3}})=x_{1}y_{1}+x_{2}y_{2}+
x_{3}y_{3}.
$$
The scalar product in the current algebra ${\cal G_{\lambda}}$ is defined by
$$
<X,\, Y>=\hbox{res} (X,\,  Y),
\qquad X,Y\in {\cal G}_{\lambda}.
$$

Evidently, any factorizing subalgebra  ${\cal G}_{+}$ contains unique elements
of the form
\begin{equation}\label{XYZ}
{\bf E_{1}}=\frac{1}{\lambda}\,{\bf e_{1}}+O(\lambda),\qquad
{\bf E_{2}}=\frac{1}{\lambda}\,{\bf e_{2}}+O(\lambda),\qquad
{\bf E_{3}}=\frac{1}{\lambda}\,{\bf e_{3}}+O(\lambda)
\end{equation}
that generate ${\cal G}_{+}$ as a Lie algebra. It is also clear that if we take arbitarary
series (\ref{XYZ}) and generate a Lie subalgebra
${\cal G}_{+}$, then the sum of this susbalgebra and the algebra of all Taylor series coincides with
the whole current algebra. The only problem is that, in general, this sum is not direct.

It follows from the invariance of the scalar product
$<\cdot,\cdot>$ that ${\cal G}_{+}^{\perp}$ is a module over ${\cal G}_{+}$.
Obviously, ${\cal G}_{+}^{\perp}$ contains unique elements of the form
\begin{equation}\label{RXYZ}
{\bf R_{1}}=\frac{1}{\lambda}\,{\bf e_{1}}+O(\lambda),\qquad
{\bf R_{2}}=\frac{1}{\lambda}\,{\bf e_{2}}+O(\lambda),\qquad
{\bf R_{3}}=\frac{1}{\lambda}\,{\bf e_{3}}+O(\lambda)
\end{equation}
that generate ${\cal G}_{+}^{\perp}$ as a module over ${\cal G}_{+}$.

{\bf Example 1.} It can  be checked that the generators
$$
{\bf E_{1}}=\frac{1}{\lambda}\,{\bf e_{1}} (1+q \lambda)^{1/2}(1+r
\lambda)^{1/2},
$$
$$
{\bf E_{2}}=\frac{1}{\lambda}\,{\bf e_{2}} (1+p \lambda)^{1/2}(1+r
\lambda)^{1/2},
$$
$$
{\bf E_{3}}=\frac{1}{\lambda}\,{\bf e_{3}} (1+p \lambda)^{1/2}(1+q \lambda)^{1/2}
$$
defines a factorizing subalgebra (see \cite{golsok2}). The orthogonal
complement is generated by
$$
{\bf R_{1}}=\frac{1}{\lambda}\,{\bf e_{1}} (1+q \lambda)^{-1/2}(1+r
\lambda)^{-1/2},
$$
$$
{\bf R_{2}}=\frac{1}{\lambda}\,{\bf e_{2}} (1+p \lambda)^{-1/2}(1+r
\lambda)^{-1/2},
$$
$$
{\bf R_{3}}=\frac{1}{\lambda}\,{\bf e_{3}}
(1+p \lambda)^{-1/2}(1+q \lambda)^{-1/2}.
$$
The simplest orbit is given by (\ref{orbnst}) with $n=1$. It consists of the operators of the form
$$
L=M_{1}{\bf R_{1}}+M_{2}{\bf R_{2}}+M_{3}{\bf R_{3}}.
$$
The function $H_{\lambda}=(L,\,L)$ is a polynomial in $\mu=\lambda^{-1},$
whose coefficients are functions
$$
H_{1}=\hbox{res}(\lambda H_{\lambda})=M_{1}^{2}+M_{2}^{2}+M_{3}^{2}
$$
and
$$
H_{2}=\hbox{res}(H_{\lambda})=p M_{1}^{2}+q M_{2}^{2}+r M_{3}^{2}.
$$
Since in this case the algebra (\ref{algebra1}) is isomorphic to $so(3)$,
these polynomials commute each other with respect to the Poisson bracket
\begin{equation} \label{puas1}
\{M_{i},M_{j}\}=\varepsilon_{ijk}\,M_{k},
\end{equation}
where $\varepsilon_{ijk}$ is the totally skew-symmetric tensor.

Taking for $A$ the operator
$$
A=\pi_{+}(L)=M_{1}{\bf E_{1}}+M_{2}{\bf E_{2}}+M_{3}{\bf E_{3}},
$$
related to the Hamiltonian $H_{2}$ by (\ref{gf}), we get a Lax pair for the
Euler top
$$
{\bf M}_{t}+{\bf M}\times V {\bf M}=0,
$$
where
\begin{equation}\label{MatV}
V=\pmatrix{p&0&0 \cr 0&q&0 \cr 0&0&r\cr}.
\end{equation}

The orbit (\ref{orbnst}) with $n=2$ consists of the operators of the form
\begin{equation}\label{Lax2}
L=\gamma_{1}{\bf S_{1}}+\gamma_{2}{\bf S_{2}}+\gamma_{3}{\bf S_{3}}+
M_{1}{\bf R_{1}}+M_{2}{\bf R_{2}}+M_{3}{\bf R_{3}}.
\end{equation}
If we choose the adjoint bases in ${\cal O}$ and ${\cal H}$ (see section 2),
then
$$
{\bf S_{1}}= [{\bf R_{3}},\, {\bf E_{2}}  ]+\frac{1}{2}(q-r){\bf R_{1}}, \qquad
{\bf S_{2}}= [{\bf R_{1}},\, {\bf E_{3}}  ]+\frac{1}{2}(r-p){\bf R_{2}}, \qquad
{\bf S_{3}}= [{\bf R_{2}},\, {\bf E_{1}}  ]+\frac{1}{2}(p-q){\bf R_{3}}.
$$
The function $H_{\lambda}=(L,\,L)$ is (up to a multiplier) a polynomial in
$\mu=\lambda^{-1}.$ As algbera
(\ref{algebra1}) is isomorphic to $e(3),$ the coefficients of this
polynomial commute each other with respect to the Poisson bracket
\begin{equation} \label{puas}
\{M_{i},M_{j}\}=\varepsilon_{ijk}\,M_{k}, \qquad
\{M_{i},\gamma_{j}\}=\varepsilon_{ijk}\,\gamma_{k}, \qquad
\{\gamma_{i},\gamma_{j}\}=0.
\end{equation}
Two non-trivial coefficients are as follows
$$
H_{1}=M_{1}^{2}+M_{2}^{2}+M_{3}^{2}+ p M_{1} \gamma_{1}+q M_{2} \gamma_{2}+
r M_{3} \gamma_{3}+\frac{1}{4}(q-r)^{2} \gamma_{1}^{2}+
\frac{1}{4}(p-r)^{2} \gamma_{2}^{2}+\frac{1}{4}(p-q)^{2} \gamma_{3}^{2}
$$
and
$$
H_{2}=p M_{1}^{2}+q M_{2}^{2}+r M_{3}^{2}
+p(q+r) M_{1} \gamma_{1}+q(p+r) M_{2} \gamma_{2}+
r(p+q) M_{3} \gamma_{3}
$$
$$+\frac{1}{4}p(q+r)^{2} \gamma_{1}^{2}+
\frac{1}{4}q(p+r)^{2} \gamma_{2}^{2}+\frac{1}{4}r(p+q)^{2} \gamma_{3}^{2}.
$$
This commuting pair of quadratic polynomials defines the integrable Steklov-Lyapunov
case in the Kirchhoff problem of the motion of a
rigid body in an ideal fluid. The equations of motion defined by the
Hamiltonian $H_{1}$ correspond to $A$-operator of the form
$$
A=\pi_{+}(\lambda L)=M_{1}{\bf E_{1}}+M_{2}{\bf E_{2}}+M_{3}{\bf E_{3}}.
$$
The flow defined by the Hamiltonian $H_{2}$ corresponds to $A=\pi_{+}(L)$.

{\bf Example 2.} One can verify that the generators
$$
{\bf E_{1}}=\frac{1}{\lambda}\,{\bf e_{1}} \sqrt{1-p \lambda^{2}}, \qquad
{\bf E_{2}}=\frac{1}{\lambda}\,{\bf e_{2}} \sqrt{1-q \lambda^{2}}, \qquad
{\bf E_{3}}=\frac{1}{\lambda}\,{\bf e_{3}} \sqrt{1-r \lambda^{2}}
$$
give rise to a factorizing subalgebra. This subalgebra is isotropic with
respect to the form
$<\cdot,\cdot>$ and therefore ${\bf R_{i}}={\bf E_{i}}$.
The orbit (\ref{orbnst}) with $n=2$ consists of operators of the form (\ref{Lax2}),
where
$$
{\bf S_{1}}= [{\bf E_{2}},\,{\bf E_{3}}] \qquad
{\bf S_{2}}= [{\bf E_{3}},\,{\bf E_{1}}], \qquad
{\bf S_{3}}= [{\bf E_{1}},\,{\bf E_{2}}].
$$
The function $H_{\lambda}=(L,\,L)$ leads to two non-trivial quadratic
polynomial
$$
H_{1}=M_{1}^{2}+M_{2}^{2}+M_{3}^{2}-(q+r) \gamma_{1}^{2}-
(p+r) \gamma_{2}^{2}-(p+q) \gamma_{3}^{2}
$$
and
$$
H_{2}=p M_{1}^{2}+q M_{2}^{2}+r M_{3}^{2}
-q r\, \gamma_{1}^{2}-p r\, \gamma_{2}^{2}-p q\, \gamma_{3}^{2}.
$$
This pair of quadratic polynomials commuting with respect to bracket (\ref{puas})
defines the  Clebsch
integrable case in the Kirchhoff problem of the motion of a
rigid body in an ideal fluid. Equations of motions for the Hamiltonian $H_{1}$ corresponds to operator
$A=\pi_{+}(\lambda L)$. The flow for $H_{2}$ is given by $A=\pi_{+}(L)$.

{\bf Example 3.} Let
\begin{equation}\label{RRR}
R=\pmatrix{0&r_{3}&r_{1} \cr -r_{3}&0&r_{2} \cr -r_{1}&-r_{2}&0
\cr}
\end{equation}
be arbitrary constant element of $so(3)$. Then the generators
$$
{\bf E_{1}}=\frac{{\bf e_{1}}}{\lambda}+\nu\, [R,{\bf e_{1}}]+
\frac{1}{2}[R,\,[R,{\bf e_{1}}]],
$$
$$
{\bf E_{2}}=\frac{{\bf e_{2}}}{\lambda}+\nu\, [R,{\bf e_{2}}]+
\frac{1}{2}[R,\,[R,{\bf e_{2}}]],
$$
$$
{\bf E_{3}}=\frac{{\bf e_{3}}}{\lambda}+\nu\, [R,{\bf e_{3}}]+
\frac{1}{2}[R,\,[R,{\bf e_{3}}]]
$$
define a factorizing subalgebra. The generators of the orthogonal complement are given by
$$
{\bf R_{1}}=\frac{{\bf e_{1}}}{\lambda}+\nu\, [R,{\bf e_{1}}]-
\frac{1}{2}[R,\,[R,{\bf e_{1}}]]-(R,\,R){\bf e_{1}},
$$
$$
{\bf R_{2}}=\frac{{\bf e_{2}}}{\lambda}+\nu\, [R,{\bf e_{2}}]-
\frac{1}{2}[R,\,[R,{\bf e_{2}}]]-(R,\,R){\bf e_{2}},
$$
$$
{\bf R_{3}}=\frac{{\bf e_{3}}}{\lambda}+\nu\, [R,{\bf e_{3}}]-
\frac{1}{2}[R,\,[R,{\bf e_{3}}]]-(R,\,R){\bf e_{3}}.
$$
The orbit with $n=2$ is described by formula (\ref{Lax2}), where
$$
{\bf S_{1}}= [{\bf E_{2}},\,{\bf R_{3}}]+
\frac{1}{2}(r_{2}^{2}-r_{3}^{2})\,{\bf R_{1}}, \quad
{\bf S_{2}}=  [{\bf E_{3}},\,{\bf R_{1}}]+
\frac{1}{2}(r_{3}^{2}-r_{1}^{2})\,{\bf R_{2}},, \quad
{\bf S_{3}}=  [{\bf E_{1}},\,{\bf R_{2}}]+
\frac{1}{2}(r_{1}^{2}-r_{2}^{2})\,{\bf R_{3}}.
$$
Coefficients of $H_{\lambda}=(L,\,L)$ yield two non-trivial quadratic polynomials
$H_{1}$ and  $H_{2}$, which commute with respect to the Poisson bracket (\ref{puas}).
It is not difficult to find a linear combination of $H_{1}$ and the Casimir functions for
bracket (\ref{puas}) that equals the square of the following linear integral
$$
I=r_{1}M_{1}+ r_{2} M_{2}+r_{3}M_{3}+\frac{1}{2}r_{1}(r_{2}^{2}-r_{3}^{2})
\gamma_{1}+\frac{1}{2}r_{2}(r_{3}^{2}-r_{1}^{2})
\gamma_{2}+\frac{1}{2}r_{3}(r_{1}^{2}-r_{2}^{2})
\gamma_{3}.
$$
Thus the factorizing subalgebra from Example 3 gives rise to a Lax
representation for the Kirchhoff integrable case in the problem of the motion of a
rigid body in an ideal fluid.

The factorizing subalgebras from Examples 2,3 are invariant with respect to
multiplying by $\lambda^{-2},$ whereas the subalgebra from Example 1
admits multiplication by $\lambda^{-1}.$
According to remark \ref{rem3}, the corresponding flows are hamiltonian with respect to one more
linear bracket defined by the Lie algebra
$
{\cal H}={\cal G}_{+}/ \lambda^{-2} {\cal G}_{+}
$,
which is isomorphic to $so(4)$ for generic values of parameters. This fact explains the
existence of linear transformations found by A. Bobenko \cite{bobenko}, which
interconnect known integrable cases on $e(3)$ and $so(4)$.

As it was shown in \cite{golsok1}, any factorizing subalgebra for $so(3)$ yields an integrable
PDE of the Landau-Lifshitz type for one vector unknown
 ${\bf s}=(s_{1},s_{2},s_{3})$, where $\vert {\bf s} \vert=1$.
In this case the $L$-operator is of the form
\begin{equation}
L=s_{1} {\bf E_{1}}+s_{2} {\bf E_{2}}+s_{3} {\bf E_{3}},
\end{equation}
and the $A$-operator has the following structure:
\begin{equation}
A=s_{1} {\bf E_{2}}\times {\bf E_{3}}+s_{2} {\bf E_{3}}\times {\bf E_{1}}+
s_{3} {\bf E_{1}}\times {\bf E_{2}}
+q_{1} {\bf E_{1}}+q_{2} {\bf E_{2}}+q_{3} {\bf E_{3}},
\end{equation}
where $q_{i}$ are some (individual for each factorizing subalgebra)
differential polynomials in the components of the vector ${\bf s}.$ Their explicit form can be
easily determined from the Lax equation (\ref{laxpde}).

It turns out that for Example 2 relation
(\ref{laxpde}) is equivalent (up to change of signs for independent variables) to the
Landau-Lifshitz equation
$$
{\bf s}_{t}={\bf s}\times {\bf s}_{xx}+ {\bf s} \times V{\bf s},
$$
where the matrix $V$ is given by (\ref{MatV}). For the first time, this Lax
representation for the Landau-Lifshitz equation was found in \cite{sklyan}.

The subalgebra from Example 1 yields the equation
\begin{equation} \label{L-L1}
{\bf s}_{t}={\bf s}\,\times \Big({\bf s}_{xx}+\frac{1}{2}(V [{\bf s}_{x},{\bf s}]+
[{\bf s}_{x},{\bf s}] V)+\frac{1}{2}[{\bf s},\, V {\bf s}_{x}+{\bf
s}_{x}V]+[{\bf s}_{x},\, V {\bf s}+{\bf s}V]\Big).
\end{equation}

At last, the subalgebra from Example 3 corresponds to equation
$$
{\bf s}_{t}={\bf s}\,\times \Big({\bf s}_{xx}+\frac{1}{2}(Z [{\bf s}_{x},{\bf s}]+
[{\bf s}_{x},{\bf s}] Z)+\frac{1}{2}[{\bf s},\, Z {\bf s}_{x}+{\bf
s}_{x}Z]+[{\bf s}_{x},\, Z {\bf s}+{\bf s}Z]+c\, Z{\bf s}\Big),
$$
where
$$
Z=\pmatrix{r_{1}^{2}&r_{1}r_{2}&r_{1}r_{3} \cr r_{1}r_{2}&r_{2}^{2}&r_{2}r_{3}
 \cr r_{1}r_{3}&r_{2}r_{3}&r_{3}^{2}\cr}, \qquad
 c=\nu^{2}+\frac{r_{1}^{2}+r_{2}^{2}+r_{3}^{2}}{4}.
$$
In other words, one can say that $Z$ is arbitrary matrix of rank 1 and
 $c$ is arbitrary constant. The above list of three equations coincides with
 the list of the paper \cite{mikhshab}, where all equations of the Landau-Lifshitz type having higher
 conservation laws have been found.

\section{Other examples.}
\setcounter{equation}{0}
\subsection{Generalization of Example 1 to the  $so(n)$-case.}
The factorizing subalgebra from Example 1 can be defined by the formula
\begin{equation}\label{Uplus}
{\cal U}_{+}=(1+\lambda V)^{1/2}{\cal U}^{st} (1+\lambda V)^{1/2},
\end{equation}
where ${\cal U}^{st}$ denotes the set of polynomials in
$\lambda^{-1}$ with coefficients from $so(3)$, and
the matrix $V$ is given by (\ref{MatV}). It can easily be checked that
formula (\ref{Uplus}), where $V$ is arbitrary diagonal matrix,
defines a factorizing sublagebra for
${\cal G}=so(n)$ as well.
The orthogonal complement to ${\cal U}_{+}$ is given by
$$
{\cal U}_{+}^{\perp}=(1+\lambda V)^{-1/2}{\cal U}^{st} (1+\lambda V)^{-1/2}.
$$
The simplest orbit (\ref{orbnst}) corresponding to $n=1$ gives rise to the following hamiltonian
equation
$$
{\bf M}_{t}=[V, \, {\bf M}^{2}], \qquad {\bf M}\in so(n)
$$
on $so(n)$ with the Lax pair
$$
L=(1+\lambda V)^{-1/2}\,\frac{{\bf M}}{\lambda}\, (1+\lambda V)^{-1/2},
\qquad A=(1+\lambda V)^{1/2}\,\frac{{\bf M}}{\lambda}\, (1+\lambda V)^{1/2}.
$$
The equation
$$
{\bf M}_{t}=[V, \, {\bf M}^{2}]+[{\bf M},\,{\bf \Gamma}],
\qquad {\bf \Gamma}_{t}=V {\bf M}  {\bf \Gamma} -{\bf \Gamma} {\bf M} V,
\qquad {\bf M},{\bf \Gamma}\in so(n)
$$
corresponding to the orbit with $n=2,$ possesses the Lax pair
$$
L=(1+\lambda V)^{-1/2}\,\left(\frac{{\bf M}}{\lambda^{2}}+
\frac{{\bf \Gamma}}{\lambda}\right)\, (1+\lambda V)^{-1/2},
\qquad A=(1+\lambda V)^{1/2}\,\frac{{\bf M}}{\lambda}\, (1+\lambda V)^{1/2}.
$$
One can regard this equation as a  $so(n)$-generalization of the Steklov-Lyapunov top.
In this case the dynamical variables are formed in a pair of matrices from $so(n)$ and the linear
Poisson bracket is defined by the Lie algebra
$so(n)\oplus \epsilon so(n),$ where $\epsilon^{2}=0.$

For the generalized  Landau-Lifshitz equation
$$
{\bf S}_{t}={\bf P}_{x}+{\bf P}V{\bf S}-{\bf S}V{\bf P}, \qquad
{\bf S}_{x}=[{\bf S},\,{\bf P}],
$$
related to subalgebra (\ref{Uplus}), the Lax representation is given by
$$
L=(1+\lambda V)^{1/2}\,\frac{{\bf S}}{\lambda}\, (1+\lambda V)^{1/2}, \qquad
A=(1+\lambda V)^{1/2}\,\left(\frac{{\bf S}}{\lambda^{2}}+
\frac{{\bf P}}{\lambda}\right)\, (1+\lambda V)^{1/2}.
$$
For the Lie algebra $so(3)$ on the orbit $({\bf S},\, {\bf S})=1,$ we have
${\bf P}=[{\bf S},\,{\bf S}_{x}]+q {\bf S}, $ where the unknown function $q$
has to be found from the condition $({\bf S},\, {\bf S}_{t})=0$. As the result, we get
(up to $t\rightarrow -t$) equation (\ref{L-L1}).

\subsection{An example on the Kac-Moody algebra related to $so(n,m)$.}

Let
$${\cal G}=\{A\in Mat_{(n+m)\times (n+m)}\, \vert\, A^{T}=-S A S \},$$
where
$$
S=\left(\begin{array}{cc}
  E_{n} & 0 \\
  0 & -E_{m} \\
\end{array}\right).
$$
It is clear that the Lie algebra ${\cal G}$ is isomorphic to  $so(n+m).$
Consider the subalgebra ${\cal U}$ of the current algebra over ${\cal G}$ consisting of such Laurent series
that the coefficients of even (respectively, odd) powers of $\lambda$  belong to $V_{1}$
(respectively, $V_{-1}$). Here by $V_{\pm 1}$ denote eigenspaces of the inner second
order automorphism  ${\cal G}\rightarrow S{\cal G}S^{-1},$ corresponding to eigenvalues  $\pm 1$.
Actually, this means that the coefficients of even powers of
$\lambda$ have the following block structure
$$
\left(\begin{array}{cc}
  v_{1} & 0 \\
  0 & v_{2} \\
\end{array}\right),
$$ where $v_1\in so(n), \, v_2\in so(m)$, and the coefficients of odd powers
are of the form
$$ \left(\begin{array}{cc}
  0 & w \\
  w^t & 0 \\
\end{array}\right),
$$
where $w\in Mat_{n,m}$.

We choose $\hbox{res} (\lambda^{-1}\hbox {tr}(X\,Y))$ for the non-degenerate invariant form
on ${\cal U}$. Note that in this case the form $\hbox{res} (\hbox {tr}(X\,Y))$ is generate.

Let ${\cal U}_{-}$ be the set of all Taylor series from ${\cal U}$,
$$
{\cal U}_{+}=(1+\lambda r)^{1/2}{\cal U}^{st} (1+\lambda r)^{1/2},
$$
where ${\cal U}^{st}$ is the set of polynomials in
$\lambda^{-1}$ from ${\cal U}$ and $r$ is arbitrary constant matrix of the form
$$
r=\left(\begin{array}{cc}
  0 & r_1 \\
  -r_1^t & 0 \\
\end{array}\right).
$$
Such choice of factorizing subalgebra is a natural generalization of Example \ref{rrr} to the
case, when the structure of coefficients of series from ${\cal U}$ are
defined by an additional automorphism of second order.

Suppose $I_{-}=\lambda^{2} {\cal U}_{-}$ and $$I_{+}=\{(1+\lambda
r)^{1/2} \sum_{i=2}^{k} q_i \lambda^{-i}(1+\lambda r)^{1/2} \}.$$
Then the orbit (\ref{genort}) consists of the elements of the form
\begin{equation}\label{koworb}
 L= (1+\lambda
r)^{-1/2}(\lambda^{-1} w+v+\lambda u)(1+\lambda r)^{-1/2}.
\end{equation}
The corresponding Lax equation
\begin{equation}\label{genlax}
L_{t}=[\pi_{+}(L),\, L]
\end{equation}
is equivalent to the system of equations
\begin{equation}\label{sys}
w_{t}=[w,\, wr+rw-v], \quad v_{t}=[u,\,w]+vwr-rwv, \quad u_{t}=uwr-rwu.
\end{equation}
It is easy to see that this system admits the reduction
$$
u=\left(\begin{array}{cc}
  0 & r_1 \\
  r_1^t & 0 \\
\end{array}\right),
$$
which is equivalent to the model found in \cite{sokts1}.
In the case $r=0$ the system and it's Lax representation have been considered in \cite{brs89}.

Let us consider the case  ${\cal G}=so(3,2)$ in more detail. Without loss of
generality the matrices $u$ and $r$ may be chosen as follows
\begin{equation}\label{uu}
u=\left(\begin{array}{ccccc}
   0 & 0 & 0 &a_1 &0 \\
   0 & 0 & 0 &0 &a_{2} \\
   0 & 0 & 0 &0 &0 \\
   a_{1} & 0 & 0 &0 &0 \\
   0 & a_{2} & 0 &0 &0 \\
\end{array}\right), \qquad
r=\left(\begin{array}{ccccc}
   0 & 0 & 0 &k_1 &0 \\
   0 & 0 & 0 &0 &k_{2} \\
   0 & 0 & 0 &0 &0 \\
   -k_{1} & 0 & 0 &0 &0 \\
   0 & -k_{2} & 0 &0 &0 \\
\end{array}\right),
\end{equation}
where $a_{1} k_{2}=a_{2} k_{1}.$ Consider a special case
$k_{2}=k_{1}=k, \,a_{2}=a_{1}=a$, which is remarkable since the
corresponding Hamiltonian admits an additional hamiltonian reduction (see \cite{reysem}).
In this case the $L$-operator is given by
(\ref{koworb}), where
\begin{equation}\label{ww}
w=\left(\begin{array}{ccccc}
   0 & 0 & 0 &\gamma_{1} &\delta_{1} \\
   0 & 0 & 0 &\gamma_{2} &\delta_{2} \\
   0 & 0 & 0 &\gamma_{3} &\delta_{3} \\
   \gamma_{1} & \gamma_{2} & \gamma_{3} &0 &0 \\
   \delta_{1} & \delta_{2} & \delta_{3} &0 &0 \\
\end{array}\right)
\end{equation}
and, as a simple computation shows,
$$
v=\left(\begin{array}{ccccc}
   0 & -M_{3} & M_{2} &0 &0 \\
   M_{3} & 0 & -M_{1} &0 &0 \\
   -M_{2} & M_{1} & 0 &0 &0 \\
   0 & 0 & 0 &0 &M_{4}+k(\gamma_{2}-\delta_{1}) \\
   0 & 0 & 0 &-M_{4}-k(\gamma_{2}-\delta_{1}) &0 \\
\end{array}\right).
$$
Among the spectral invariants, there is the integral $(M_{4}-M_{3})^{2}$.
Reduction mentioned above is that we fix the integral value
: $M_{4}-M_{3}=z$. Note that in order to the dynamics defined by Lax equation (\ref{laxode}) to
be consistent with the reduction, an appropriate skew-symmetric matrix has to be added to the operator
$A= \pi_{+}(L)$ (cf. \cite{reysem}).

One can check that under the reduction $\hbox{tr} L^{2}$ yields the
Hamiltonian (see \cite{sokts1})
$$
H = M_{1}^2 + M_{2}^2 + 2 M_{3}^2  + 2 k (M_{3} \gamma_{2} - M_{2} \gamma_{3})
+ 2 k (M_{1} \delta_{3}- M_{3} \delta_{1}) - 2 a \gamma_{1}-
2 a \delta_{2}  + 2 z M_{3},
$$
while $\hbox{tr} L^{4}$ gives us two fourth degree integrals providing the Liouville integrability
of this Hamiltonian.  The further reduction
$\delta_{1}=\delta_{2}=\delta_{3}=0$ leads to the integrable case for the  Kirchhoff problem found in
 \cite{soktmf}. The Poisson bracket for this reduction corresponds to the Lie algebra $e(3)$
(see formula (\ref{puas})).

One of possible integrable generalizations for the reduced Hamiltonian to the family of brackets
\begin{equation} \label{puaso4}
\{M_{i},M_{j}\}=\varepsilon_{ijk}\,M_{k}, \qquad
\{M_{i},\gamma_{j}\}=\varepsilon_{ijk}\,\gamma_{k}, \qquad
\{\gamma_{i},\gamma_{j}\}=\kappa \varepsilon_{ijk}\,M_{k}.
\end{equation}
has been proposed in \cite{bomasok}. It turns out that
\begin{equation}\label{newkoworb}
 L= R^{-1/2} (\lambda^{-1} w+v+\lambda u) R^{-1/2}
\end{equation}
with
$$
R=\left(\begin{array}{ccccc}
   1 & 0 & 0 &k_1 \lambda &0 \\
   0 & 1 & 0 &0 &k_{2} \lambda\\
   0 & 0 & 1 &0 &0 \\
   -k_{1} \lambda & 0 & 0 &1-\kappa\lambda^{-2} &0 \\
   0 & -k_{2} \lambda& 0 &0 &1 \\
\end{array}\right),
$$
can be taken for a Lax operator for this model.
The parameter $\kappa$ from the Poisson bracket (\ref{puaso4}) is related to $k_1, k_2$
by the formula
$$\kappa=\frac{k_1^2-k_2^2}{k_2^4}.$$
The matrix $u$ is defined by (\ref{uu}), where $a_{1} k_{2}=a_{2} k_{1}.$
The matrix $w$ is given by (\ref{ww}) with $\delta_1=\delta_2=
\delta_3=0.$ At last, the matrix $v$ is of the form
$$
v=\left(\begin{array}{ccccc}
   0 & -M_{3} & M_{2} &0 &0 \\
   M_{3} & 0 & -M_{1} &0 &0 \\
   -M_{2} & M_{1} & 0 &0 &0 \\
   0 & 0 & 0 &0 &\tau M_{3}+k_2 \gamma_{2}+z \\
   0 & 0 & 0 &-\tau M_{3}-k_2 \gamma_{2}-z &0 \\
\end{array}\right),
$$
where $\tau=\frac{k_1}{k_2}$. The spectral invariants of the operator $L$
commute each other with respect to bracket (\ref{puaso4}). The Hamiltonian
$$
H=\tau^2 M_1^2+M_2^2+2 \tau^2 M_3^2+2 k_1 (M_3 \gamma_2-M_2 \gamma_3)
-2 a_1 \gamma_1+2 \tau z M_3
$$
is one of them. If $\tau=1,$ the Poisson bracket corresponds to the algebra $e(3)$ and
the $L$-operator coincides with the one described above. In general case,
the algebraic nature of the matrix $R$ remains to be mysterious
(cf. \cite{kst}).

Explicit formulas for other integrable deformations of models from the paper
 \cite{brs89} can be found in \cite{sokts1,ts}. However the Lax
 representations presented in these papers are not as algebraic and elegant as (\ref{koworb}).
 In spite of that they looks similar to (\ref{koworb}), the matrix $r$ is variable.

\section{Homogeneous subalgebras.}
\setcounter{equation}{0}

To efficiently describe the finite-dimensional orbits corresponding a
factorizing subalgebra ${\cal G}_{+}$, it is necessary to find
the vector space ${\cal G}_{+}^{\perp}$ in an explicit form. It
easily can be done in the case, when ${\cal G}_{+}$ is a homogeneous subalgebra (see \cite{golsok2}).

A factorizing subalgebra ${\cal G}_{+}$ is called {\bf homogeneous} if the following condition holds
\begin{equation}
\frac{1}{\lambda} \ {\cal G}_{+} \subset {\cal G}_{+}. \label{cond}
\end{equation}

Examples of homogeneous subalgebras can be found in
\cite{borbol,golsok2}.

Let $A(\lambda)$ be a formal series of the form
\begin{equation}
A=E+ R \ \lambda+ S \ \lambda^2 + \cdots \label{RS}
\end{equation}
where $R, S, \dots$ are linear constant operators from ${\cal G}$ to ${\cal G}$, $E$ is
the identity operator.

\begin{thr} (see \cite{golsok2})
\begin{itemize}
\item i) \,  Any homogeneous subalgebra ${\cal G}_{+}$ can be represented in
the form
\begin{equation}\label{Aaa}
{\cal G}_{+}=\{ \sum ^{k}_{i=1} \lambda^{-i} \ A(g_i), \ \vert \ g_i \in {\cal
G}, \ \ k\in {\mathbb N}  \} \label{AA}
\end{equation}
where $ A(\lambda)$ is a formal series of the form (\ref{RS}).
\item ii) \, Vector space (\ref{Aaa}) is a subalgebra iff
for any $X,Y \in {\cal G}$
\begin{equation}
[A(X), \ A(Y)]=A \Big( [X, \ Y]+\lambda \ [X, \ Y]_1\Big), \label{sog2}
\end{equation}
where $[\cdot,\cdot]_1$ is a Lie bracket compatible with the bracket
$[\cdot,\cdot]$
\item iii)  For any homogeneous subalgebra the bracket $[\cdot,\cdot]_1$ is given by
\begin{equation} \label{brac1}
[X, \, Y]_1=[R(X), \, Y]+[X, \, R(Y)]-R([X, \, Y]),
\end{equation}
where $R$ is the corresponding coefficient in (\ref{RS}).
\end{itemize}
\end{thr}
Recall that the compatibility of two Lie brackets means that arbitrary
linear combination of these brackets is also a Lie bracket.

For all homogeneous subalgebras ${\cal G}_{+}$ the orthogonal complement
${\cal G}_{+}^{\perp}$ can be found my means of the same formula.

\begin{thr} Let ${\cal G}_{+}$ be a homogeneous subalgebra. Then
\begin{equation}\label{orthom}
{\cal G}_{+}^{\perp}=(A^{-1})^{T} ({\cal G}_{+}^{st}),
\end{equation}
where the series $A$ generates  ${\cal G}_{+}$ by formula (\ref{AA}), the standard
factorizing subalgebra ${\cal G}_{+}^{st}$ is defined by (\ref{polyn}), and
$T$ stands for transposition with respect to the scalar product (\ref{formass}).
\end{thr}

\par\noindent
{\bf Proof.} It follows from (\ref{AA}) that ${\cal G}_{+}=A({\cal
G}_{+}^{st})$. Hence
$$
<A({\cal G}_{+}^{st}), \, (A^{-1})^{T} ({\cal G}_{+}^{st}) >= <{\cal
G}_{+}^{st},\, {\cal G}_{+}^{st}>=0,
$$
i.e. ${\cal G}_{+}^{\perp}\supset (A^{-1})^{T} ({\cal G}_{+}^{st})$. From other
hand, if $g\in {\cal G}_{+}^{\perp}$, then $<A({\cal G}_{+}^{st}),\, g>=0$.
This implies
$A^{T}(g)\in ({\cal G}_{+}^{st})^{\perp}= {\cal G}_{+}^{st}$ or $g\in
(A^{-1})^{T} ({\cal G}_{+}^{st})$.

\subsection{The orbits for the case of homogeneous subalgebras.}

The following statement is convenient for finding of explicit form of Lax
equations corresponding to homogeneous subalgebras.

\begin{prop}
Suppose $A$ satisfies (\ref{sog2}); then
the following identity holds
\begin{equation}\label{iden}
[A(X),\, (A^{-1})^{T}(Y)]=(A^{-1})^{T}\left([X,Y]+\lambda X*Y \right).
\end{equation}
Here for any $X,Y \in {\cal G}$
\begin{equation}\label{star}
X*Y=[R(X),\, Y]-[X,\,R^{t}(Y)]+R^{t}([X,Y]),
\end{equation}
where $R$ is the coefficient from (\ref{RS}), $t$ stands for transposition with respect to
the invariant form on  ${\cal G}$.
\end{prop}

To prove identity (\ref{iden}) it is suffice to multiply innerly both sides by
$A(Z)$ and transform the left hand side using the invariance of the scalar product
$<\cdot,\cdot>$ and identity (\ref{sog2}).

Notice that
$$
X*Y=-({\rm ad}_{1} X)^{t} (Y),
$$
where ${\rm ad}_{1} X (Y)=[X,\,Y]_{1}$ and the bracket $[\cdot,\cdot]_{1}$ is given
by (\ref{brac1}). In other words, the operator $-({\rm ad}_{1} X)^{t}$
defines the coadjoint action of the Lie algebra equipped with the bracket
$[\cdot,\cdot]_{1}$ on the Lie algebra with the bracket $[\cdot,\cdot]$.

Any element of the orbit
$$
{\cal O}=\lambda^{-1}{\cal G}_{-}\cap \lambda^{2}{\cal G}_{+}^{\perp},
$$
corresponding to the homogeneous subalgebra (\ref{AA}) has the form
$$
L=(A^{-1})^{T}(\lambda u+v+\lambda^{-1} w).
$$
It is easy to see that $\pi_{+}(L)=A(\lambda^{-1} w)$.

Using identity (\ref{iden}), one can obtain that the component-wise form of the Lax equation
(\ref{genlax}) is
\begin{equation}\label{lax21}
w_{t}=[w,\,v]+w*w, \qquad v_{t}=[w,\,u]+w*v, \qquad u_{t}=w*u,
\end{equation}

{\bf Continuation of Example \ref{rrr} }. In the situation of Example \ref{rrr},
$A$ is the operator of right
multiplication by  $1+\lambda r$, $(A^{-1})^{T}$ is the operator of left
multiplication by $(1+\lambda r)^{-1}$ and an element of the orbit is given by
\begin{equation}\label{LL}
L=(1+\lambda r)^{-1} (\lambda u+v+\lambda^{-1} w).
\end{equation}
Formula (\ref{star}) reads as follows $X*Y=r X Y-Y X r$.
The Lax equation (\ref{genlax}) written in components is equivalent to
\begin{equation}\label{sys1}
w_{t}=[w,\, wr+rw-v], \quad v_{t}=[u,\,w]+vwr-rwv, \quad u_{t}=uwr-rwu.
\end{equation}

Note that the Lax equation is equivalent to the following Lax triad
\begin{equation}\label{triLax}
P_{t}=P \lambda^{-1}w (1+\lambda r)-(1+\lambda r)  \lambda^{-1}w P,
\end{equation}
where $P=\lambda u+v+\lambda^{-1} w$.

Suppose that the algebra $A$ is equipped with an involution $*$ \footnote {i.e.
$(a b)^{*}=b^{*}a^{*}$ and $(a^{*})^{*}=a$}. Then system (\ref{sys1}) admits two following reductions:
\begin{itemize}
\item $r$ is symmetric and $u,v,w$ are skew-symmetric with respect to $*$;
\item $r$ and $v$ are skew-symmetric,  $u$ and $w$ are symmetric.
\end{itemize}
As an example of such a reduction, one can consider the case $u,v,w \in so(3)$,
$r=diag(r_{1},r_{2},r_{3})$.

Another class of reductions for system (\ref{sys1}) is determined by the condition $u=const$.
It follows from (\ref{sys1}) that in this case the idenity $uwr-rwu=0$ should be valid for all $w$.
The trivial reduction $u=const\, r$ can be reduced to $u=0$ by a shift of $v$. However there exist also
non-trivial martix reductions.

Suppose the elemets $r,$ $w$ and $v$ are of the following block structure:
$$
r=\left(\begin{array}{cc}
  0 & r_{1} \\
  r_{2} & 0 \\
\end{array}\right), \qquad w=\left(\begin{array}{cc}
  0 & w_{1} \\
  w_{2} & 0 \\
\end{array}\right), \qquad v=\left(\begin{array}{cc}
  v_{1} & 0 \\
  0 & v_{2} \\
\end{array}\right).
$$
Then choosing
$$
u=const \left(\begin{array}{cc}
  0 & r_{1} \\
  -r_{2} & 0 \\
\end{array}\right),
$$
we get self-consistent reduction of (\ref{sys1}). A further reduction
$$
v_{1}^{t}=-v_{1}, \qquad v_{2}^{t}=-v_{2}, \qquad w_{2}=w_{1}^{t}, \qquad
r_{2}=-r_{1}^{t}
$$
is available. The latter reduction and corresponding hamiltonian structures are
described in subsection 5.2.

\subsection{Groups of reduction and homogeneous subalgebras.}
One can construct examples of homogeneous subalgebras starting with so
called groups of reduction \cite{mikhailov1}. Let $T$ be a finite subgroup in the group of
linear-fractional transformations of a parameter $\lambda$. Without loss of generality we assume
that neither element of $T$ except the unity preserves the point $\lambda=\infty$.

Let us consider the vector subspace $\bar {\cal G}$ in the current algebra ${\cal
G}_{\lambda}$ consisting of elements of the form
\begin{equation}\label{aa}
a=\sum_{n=0}^{m}\sum_{t\in T} q_{n,t}\, t(\lambda^{-n}) \vert q_{n,t}\in {\cal
G},\, m\in N .
\end{equation}
We define ${\cal G}_{+}$ as the set of all  elements in $\bar {\cal
G}$ that are stable with respect to the group action of $T$. It is clear that the set
of all Laurent expansions of elements from ${\cal G}_{+}$ at the point $\lambda=0$ forms
a subalgebra in ${\cal G}_{\lambda}.$ We denote this subalgebra by ${\cal G}_{+}$ as well.

\begin{prop}
Suppose $T$ acts on the Lie algebra ${\cal G}$ without stable non-zero elements. Then
${\cal G}_{+}$ is a factorizing subalgebra  in the current algebra ${\cal G}_{\lambda}$
complementary for the subalgebra of Taylor series ${\cal G}_{-}$.
\end{prop}
\par\noindent
{\bf Proof.} Consider the element
$$
Z_q=\sum_{t\in T} t(q \lambda^{-n}).
$$
It is clear that it belongs to ${\cal G}_{+}$ and has a Laurent expansion of the form $Z_q=q
\lambda^{-n}+o(1)$. Thus ${\cal G}_{+}$ contains a series with arbitrary principle part and
therefore ${\cal G}_{\lambda}={\cal G}_{+} + {\cal G}_{-}$. It remains to prove
that ${\cal G}_{+} \cap {\cal G}_{-}=\{0\}.$ Suppose $a\in {\cal G}_{+} \cap {\cal
G}_{-}$ and $a$ contains a term of the form $q \,t(\lambda^{-n}), \quad q\ne 0, n>0$
in the decomposition (\ref{aa}). Since $a\in {\cal G}_{+}$, we have $a=t^{-1}(a).$
Therefore $a$ contains the term $t^{-1}(q) \lambda^{-n}$ and for this reason does not belong to
the algebra of Taylor series ${\cal G}_{-}$. Suppose now that $a\in {\cal G}_{+} \cap {\cal G}_{-}$ and
$a$ does not depend on $\lambda$ (i.e. $a\in {\cal G}$). Then
$a$ is an element of ${\cal G}$ stable with respect to $T$-action and
by condition, $a=0.$

Define $\bar \lambda$ by the formula
\begin{equation} \label{barlam}
\bar \lambda^{-1}=\sum_{t\in T} t(\lambda^{-1}).
\end{equation}
The function $\bar \lambda$
is invariant with respect to the action of the group $T$ and therefore
\begin{equation} \label{main}
\bar \lambda^{-1} {\cal G}_{+}\subset  {\cal G}_{+}.
\end{equation}

It can easily be shown that a change of parameter $\lambda$ of the form
\begin{equation}\label{lam}
\lambda \rightarrow \lambda +k_{2} \lambda^{2}+k_{3} \lambda^{3}+\cdots
\end{equation}
takes each factorizing subalgebra to another factorizing subalgebra. Two
subalgebras connected by such a transformation can be regarded as
equivalent.

It is easy to see that $\bar \lambda=\lambda +o(\lambda )$. Thus formula
(\ref{barlam}) defines a transformation of the form (\ref{lam}). It follows
from (\ref{main}) that ${\cal G}_{+}$ is equivalent to a homogeneous
subalgebra.

\begin{ex} Consider the commutative group $T=Z_2\times
Z_2$, consisting of elements $1,t_1,t_2$ and $t_3=t_1 t_2.$
Define  $T$-action on the Lie algebra $so(3)$ with the help of conjugations by
diagonal orthogonal matrices
$$
diag(1,1,1), \qquad diag(-1,-1,1), \qquad diag(-1,1,-1), \qquad diag(1,-1,-1),
$$
Consider the following embedding of $T$ into the group of linear-fractional
transformation. Let
$$
s_1(\lambda)=-\lambda, \qquad s_2(\lambda)=\frac{1}{\lambda}, \qquad
s_3(\lambda)=-\frac{1}{\lambda}, \qquad \tau(\lambda)=\frac{a \lambda+b}{c
\lambda+d}.
$$
Then elements of $T$ are identified with transformations $t_i=\tau^{-1}
\circ s_i \circ \tau $. The condition  $t_i(\infty)\ne \infty$ is equivalent to
$a c\ne 0$ and $a^{4} \ne c^{4}$.
Generators (\ref{XYZ}) of subalgebra ${\cal G}_{+}$ are
$$
{\bf E_{1}}=\frac{{\bf e_{1}}}{\lambda}+\sum_{i=1}^{3} t_i\left(\frac{{\bf e_{1}}}{\lambda}\right), \qquad
{\bf E_{2}}=\frac{{\bf e_{2}}}{\lambda}+\sum_{i=1}^{3} t_i\left(\frac{{\bf e_{2}}}{\lambda}\right), \qquad
{\bf E_{3}}=\frac{{\bf e_{3}}}{\lambda}+
\sum_{i=1}^{3} t_i\left(\frac{{\bf e_{3}}}{\lambda}\right).
$$
Transformation (\ref{lam}) defined by
$$
\bar \lambda^{-1}=\frac{1}{\lambda}+\sum_{i=1}^{3}
t_i\left(\frac{1}{\lambda}\right)
$$
converts the subalgebra ${\cal G}_{+}$ into a homogeneous one.
\end{ex}

\setcounter{equation}{0}


\begin{thebibliography}{10}

\bibitem{chered} I.V. Cherednik,
\newblock{Functional realizations of basis representations of factorizinf Lie groups and
algebras},
\newblock{\em Func. Anal. and Appl.} {\bf 19}(3), 36--52, 1985.

\bibitem{beldrin}  A.A. Belavin,  V.G. Drinfeld,
\newblock{On solutions of the classical Yang-Baxter equation for simple Lie algebras},
\newblock{\em Func. Anal. and Appl.} {\bf 16}(3), 1-29, 1982.

\bibitem{golsok1} I. Z. Golubchik  and  V. V.  Sokolov,
\newblock{Generalized Heizenberg equations on Z-graded Lie algebras},
\newblock{\em Teoret. and Mat. Fiz,}, {\bf 120}(2),
248-255, 1999.

\bibitem{sokdan}  V. V.  Sokolov,
\newblock{On decompositions of the current algebra over $so(3)$ into a sum of two subalgebras},
\newblock{\em Dokl. RAS.}, {\bf }(), to be published


\bibitem{borbol} A.V. Bolsinov and A.V. Borisov,
\newblock{Lax representation and compatible Poisson brackets on Lie algebras},
\newblock{\em Math. Notes}, {\bf 72}(1), 11-34, 2002.

\bibitem{golsok2}
 I. Z. Golubchik  and  V. V.  Sokolov,
\newblock{Compatible Lie brackets
and integrable equations of the principle chiral model type},
\newblock{\em Func. Anal. and Appl.}, {\bf 36}(3), 172--181, 2002.

\bibitem{soktmf}
V. V.  Sokolov,
\newblock{A new integrable case for the Kirchhoff equation},
\newblock{\em Theoret. and Math. Phys.},
{\bf 129}(1), 1335--1340, 2001.

\bibitem{sok1} V.V. Sokolov,
\newblock{Generalized Kowalewski Top: new integrable cases on $e(3)$ and
$so(4)$},
\newblock{\em CRM Proceedings and Lecture Notes},
{\bf 32}, 307-313, 2002.

\bibitem{bomasok}  A. V. Borisov,  I. S. Mamaev and  V. V. Sokolov
 \newblock{A new integrable case on $so(4)$},
\newblock{\em Dokl. RAS}, {\bf 381}(5),
614--615, 2001.

\bibitem{sokts1}
V. V. Sokolov  and  A.V. Tsiganov,
\newblock{On Lax pairs for the generalized Kowalewski and Goryachev-Chaplygin tops},
\newblock{\em Theoret. and Math. Phys.},
{\bf  131}(1), 543-549, 2002.

\bibitem{bobenko} A.I. Bobenko,
\newblock{Euler equations on $so(4)$ and $e(3)$.
Isomorphysms of integrable cases},
\newblock{\em Func. Anal. and Appl.} {\bf 20}(1), 64-66, 1986.

\bibitem{mikhailov1} A.V. Mikhailov,
\newblock{The reduction problem and the inverse scattering method}
\newblock{in "Solitons", Topics in Current Physics, R. Bullough and P. Caudrey
eds.}, \newblock{\em  New York: Springer-Verlag}, {\bf 17}, 243--285, 1980.

\bibitem{reysem}  A.G. Reyman and M.A. Semenov-{T}ian-{S}hansky,
\newblock{Integrable system. Theoretically-group approach},
\newblock{\em  Izhevsk: RC Dynamics}, 2003,  351 c.

\bibitem{sklyan}  E.K. Sklyanin,
\newblock{On complete integrability of the Landau-Lifshitz equation}.
\newblock{\em LOMI preprint}, {\bf E-3}, 1979.

\bibitem{mikhshab} A.V. Mikhailov, A.B. Shabat.
\newblock{Integrable deformations of the Heisenberg model},
\newblock{\em Phys. Lett. A},  {\bf 116}, 191--194, 1986.

\bibitem{brs89}
A.I. Bobenko, A.G. Reyman and M.A. Semenov-{T}ian-{S}hansky,
\newblock{The Kowalewski top 99 years later: a Lax pair,
generalizations and explicit solutions},
\newblock{\em Commun.\,Math.\,Phys.}, {\bf 122}, 321, 1989.

\bibitem{kst}   I. V. Komarov,  V. V. Sokolov   and  A. V. Tsiganov,
\newblock{ Poisson maps and integrable deformations of Kowalevski top},
\newblock{\em J. Phys. A}, {\bf 36}, 8035--8047, 2003

\bibitem{ts} A.V. Tsiganov,
\newblock{Integrable deformations of the tops related to algebra $so(p,q)$}.
\newblock{\em Theoret. and Math. Phys.}, {\bf }, --, 2004

\end{thebibliography}
\end{document}